\documentclass[amsmath,amssymb,12pt]{article}
\usepackage{jcappub}
\pdfoutput=1
\usepackage[utf8]{inputenc}
\usepackage{amsfonts}
\usepackage{graphicx}
\usepackage{slashed}
\usepackage{xcolor}
\usepackage{subfig}

\def	\beq	{\begin{equation}}
\def	\eeq	{\end{equation}}
\def	\be	{\begin{equation}}
\def	\ee	{\end{equation}}
\def	\ba	{\begin{eqnarray}}
\def	\ea	{\end{eqnarray}}
\def    \nn {\nonumber}
\def	\bqt	{\begin{quote}}
\def	\eqt	{\end{quote}}

\def  \pa {\partial}

\newcommand*{\affmark}[1][*]{\textsuperscript{#1}}

\title{On the Constraint Structure of Vacuum Energy Sequestering}

\author{Andrew Svesko\affmark[1] and George Zahariade\affmark[1]}
\affiliation{\affmark[1]Department of Physics and Beyond: Center for Fundamental Concepts in Science\\
Arizona State University, Tempe, Arizona 85287, USA}
\emailAdd{andrew.svesko@asu.edu}
\emailAdd{george.zahariade@asu.edu}

\abstract{We carry out the Hamiltonian analysis of the {\it local vacuum energy sequestering} model  -- a manifestly local and diffeomorphism invariant extension of general relativity which has been shown to remove the radiatively unstable contribution to the vacuum energy generated by matter loops. We find that the degravitation of this UV sensitive quantity is enforced via global relations that are a consequence of the model's peculiar constraint structure. We also show that the model propagates the proper number of degrees of freedom and thus locally reduces to general relativity on-shell.}

\begin{document}

\maketitle


\section{Introduction}
\label{secintro}
\noindent

Observations indicate that the cosmological constant is nonzero \cite{Perlmutter:1998np,Baker:1999jw}, however, we have no clear argument which explains its observed scale within the framework of quantum field theory; in fact the disagreement between the theoretical and experimental values is often quoted as being more than ${60}$ orders of magnitude. More precisely, by virtue of the equivalence principle, the vacuum energy of all the fields present in the Standard Model gravitates and behaves just like a cosmological constant. However, like any UV sensitive quantity, its value is not computable since it is formally divergent. Instead it needs to be renormalized through the addition of a bare term and the resulting physical quantity measured. The problem lies in the fact that the observed value of the cosmological constant requires a very exact cancellation between the finite part of the regularized vacuum energy at every loop order in perturbation theory \cite{Martin:2012bt,Padilla:2015aaa}. This \emph{radiative instability} constitutes one of the most important open problems of modern physics and is known as the \emph{cosmological constant problem} \cite{Zeldovich:1968ehl,Wilczek:1983as,Weinberg:1988cp}. 

Recently \cite{Kaloper:2013zca,Kaloper:2014dqa,Kaloper:2014fca} it was shown that a slight modification of general relativity provides a consistent mechanism for degravitating the radiatively unstable part of the vacuum energy generated by matter loops. The basic idea is to decouple the matter vacuum energy from gravity by adding two rigid variables without any local degrees of freedom, namely, the bare cosmological constant $\Lambda$, and the bare Planck mass $\kappa$. These global degrees of freedom are interpreted as Lagrange multipliers enforcing global constraints. The mechanism, dubbed \emph{vacuum energy sequestering}, is described by a non-additive and non-local action which, in Jordan frame, reads
\beq S\hspace{-1mm}=\hspace{-1mm}\int d^{4}x\sqrt{-g}\left[\frac{\kappa^{2}}{2}R-\Lambda-\mathcal{L}_{m}(g^{\mu\nu},\Phi)\right]+\sigma\left(\frac{\Lambda}{\mu^{4}}\right)\label{action1},\eeq
where the global interaction term $\sigma$ is outside the integral, $\mu$ is a mass scale near the quantum field theory cutoff and $\Phi$ corresponds to the matter fields of the Standard Model. To see how the mechanism works we vary the above action with respect to the global variables to obtain $\int d^4x\sqrt{-g}=\sigma'/\mu^4$ and $\int d^4x\sqrt{-g}R=0$ which, when plugged into Einstein's equations leads to 
\be
\Lambda=\frac{1}{4}\langle T^\mu{}_\mu\rangle\,,
\ee
where $\langle\dots\rangle\equiv\int d^4x\sqrt{-g} (\dots)/\int d^4x\sqrt{-g}$ is a spacetime averaging operator (sometimes dubbed \emph{historical averaging}) and $T_{\mu\nu}$ is the energy-momentum tensor. With this equation, the vacuum energy is seen to drop from Einstein's equations. Indeed if 
$T_{\mu\nu}=-V_{\text{vac}}g_{\mu\nu}+\tau_{\mu\nu}$, where $V_{\text{vac}}$ is the vacuum energy and $\tau_{\mu\nu}$ describes local excitations, then
\be
\kappa^2G_{\mu\nu}=\tau_{\mu\nu}-\frac{1}{4}\langle\tau^{\rho}{}_\rho\rangle g_{\mu\nu}\,.
\ee
Note that this cancellation happens independently of the loop order in the perturbative expansion of $V_{\text{vac}}$. The price to pay for this cancellation is a non-local, residual cosmological constant
\be
\Lambda_{\text{res}}=\frac{1}{4}\langle\tau^\mu{}_\mu\rangle\,,
\ee
which is nevertheless radiatively stable.

A manifestly local and diffeomorphism invariant formulation of this mechanism was proposed in \cite{Kaloper:2015jra,Kaloper:2016yfa,DAmico:2017ngr}  by upgrading the rigid variables $\Lambda$ and $\kappa^{2}$ to local fields, and replacing the global interaction with local expressions enforcing the on-shell rigidity of $\kappa^{2}(x)$ and $\Lambda(x)$. The basic idea consists in using 4-form terms, reminiscent of unimodular gravity models \cite{Henneaux:1989zc}. The resulting action is
\ba
S&=&\int d^{4}x\sqrt{g}\left[\frac{\kappa^{2}(x)}{2}R-\Lambda(x)-\mathcal{L}_{m}(g^{\mu\nu},\Phi)\right]\nn\\
&&+\int\left[\sigma\left(\frac{\Lambda(x)}{\mu^{4}}\right)F+\hat{\sigma}\left(\frac{\kappa^{2}(x)}{M^{2}_{P}}\right)\hat{F}\right]\;,
\label{action2}
\ea
where
\beq 
F=\frac{1}{4!}F_{\mu\nu\lambda\sigma}dx^{\mu}dx^{\nu}dx^{\lambda}dx^{\sigma}\;,
\eeq
with 
$F_{\mu\nu\lambda\sigma}=4\partial_{[\mu}A_{\nu\lambda\sigma]}$, and analogous expressions hold for $\hat{F}$. These two 4-forms introduce gauge symmetries that render $\Lambda$ and $\kappa^{2}$ constant on-shell. The functions $\sigma$ and $\hat{\sigma}$ are two smooth functions which are in principle completely arbitrary\footnote{$\sigma$ and $\hat{\sigma}$ are required to be non-linear to ensure that the problem is non-degenerate.}. The field equations
\ba
&&\kappa^{2}G_{\mu\nu}=(\nabla_{\mu}\nabla_{\nu}-g_{\mu\nu}\Box)\kappa^{2}+T_{\mu\nu}-\Lambda g_{\mu\nu}\;,\nn\\
&&\frac{\sigma'}{\mu^{4}}F_{\mu\nu\lambda\sigma}=\sqrt{g}\epsilon_{\mu\nu\lambda\sigma}\;,\quad\frac{\hat{\sigma}'}{M^{2}_{P}}\hat{F}_{\mu\nu\lambda\sigma}=-\frac{1}{2}R\sqrt{g}\epsilon_{\mu\nu\lambda\sigma}\;,\nn\\
&&\frac{\sigma'}{\mu^{4}}\partial_{\mu}\Lambda=0\;,\quad\frac{\hat{\sigma}'}{M^{2}_{P}}\partial_{\mu}\kappa^{2}=0\,,\label{eqnsofmotion}
\ea
are entirely local. Again, to see how the vacuum energy degravitates in this setup, we trace out the gravitational field equations and average them over all of spacetime, which leads to
\beq
\Lambda=\frac{1}{4}\langle T^{\mu}{}_{\mu}\rangle+\Delta\Lambda\;,
\label{lambdacond1}
\eeq
where
\beq \Delta\Lambda=-\frac{\mu^{4}}{2}\frac{\kappa^{2}\hat{\sigma}'}{M^{2}_{P}\sigma'}\frac{\int\hat{F}}{\int F}\;.\label{seqconst1}\eeq
Substituting this result into the Einstein equations 
yields
\beq 
\kappa^{2}G_{\mu\nu}=\tau_{\mu\nu}-\frac{1}{4}\langle\tau^{\rho}{}_{\rho}\rangle g_{\mu\nu}-\Delta\Lambda g_{\mu\nu}\;.
\eeq
The additional $\Delta\Lambda$ contribution is completely arbitrary, but radiatively stable once $\kappa^{2}$ is fixed to be $M_P$, and determined via cosmological measurement. Key to this radiative stability is the fact that the 4-form fluxes $\int \hat{F}$, $\int F$ (defined on the boundary of spacetime\footnote{Since the spacetime $\mathcal{M}$ is assumed to have a boundary $\pa\mathcal{M}$ in this setup, we need to supplement the action with a Gibbons-Hawking-like term~\cite{Dyer:2008hb} $\int_{\pa\mathcal{M}}d^3z\sqrt{|\gamma|}\epsilon\kappa^2 \mathcal{K}$. Here $\gamma$ and $\mathcal{K}$ are the induced metric and extrinsic curvature of $\pa\mathcal{M}$ respectively, while $\epsilon=g_{\mu\nu}n^\mu n^\nu$ and $n^\mu$ is the outward pointing normal vector to the boundary.}) are purely IR quantities that are fixed by some extrinsic process and insensitive to the UV details of the matter sector theory. More precisely, the values of the three-forms $A$ and $\hat{A}$ are constrained on the {\it whole boundary of spacetime} and the model thus has non-standard boundary conditions. On the contrary, had their values only been specified on an initial 3-surface the equations of motion~\eqref{eqnsofmotion}
would have allowed one to solve for $A$ and $\hat{A}$ as a function of the metric for all times thus making $\Delta\Lambda$ dependent on the local geometry and consequently UV sensitive.

Let us stress here that neither the original sequestering model nor its local version take graviton loop corrections into account (though a further modification of (\ref{action2}) has been shown to resolve the divergence originating from these graviton loops \cite{Kaloper:2016jsd}).

In this work we aim to perform a Hamiltonian analysis of the local vacuum energy sequestering model (\ref{action2}), and uncover the constraint structure of the theory. An analysis of the constraint structure was previously given in \cite{Bufalo:2016omb}, however, we further show how the system of constraints sheds new light upon the different global relations that enforce the degravitation of the matter sector vacuum energy. In particular we show that these arise as a consequence of the non-standard boundary conditions of the theory that pick the value of the 0-mode of some of the Lagrange multipliers. These non-standard boundary conditions are final boundary conditions on the three forms $A$ and $\hat{A}$ -- at the level of the equations of motion -- and ensure the radiative stability of the residual value of $\Lambda$. Our Hamiltonian analysis is therefore different from the traditional one in \cite{Bufalo:2016omb}, as we use a non-standard ``initial'' value formulation, which we show accounts for the novel features of the sequester model. We will also verify that this theory has the same local and global degree of freedom count as Einstein gravity. 

The layout of the paper is as follows. After we develop and analyze a toy model exhibiting many of the features of the local vacuum energy sequestering model in Section \ref{sectoymodel}, we perform an ADM split and determine the Hamiltonian of action \ref{action2} in Section \ref{secHamseq}. In Section \ref{secHamseq} we also carry out the complete Hamiltonian analysis of the model where we interpret (\ref{lambdacond1}) as a peculiar kind of global constraint, and verify that, on-shell, the local sequestering model cannot be distinguished from general relativity. We summarize our results and discuss plans for future work in Section \ref{secconc}\,.

\section{Toy model}
\label{sectoymodel}

\subsection{Action and Equations of Motion}
\noindent

We begin by studying a $0+1$ dimensional toy model that mimics most of the relevant aspects of the local sequestering mechanism. Its action is given by
\ba
S\hspace{-1.8mm}&=&\hspace{-1.8mm}\int_0^{T}\hspace{-1.8mm}dt\left(\hspace{-1mm}-\frac{1}{2}m q \ddot{q}-(\lambda+F)q+\hspace{-.5mm}f\left(\frac{\lambda}{\lambda_{0}}\right)\dot{\eta}+g\left(\frac{m}{m_{0}}\right)\dot{\rho}\right)+\left[\frac{1}{2}m q\dot{q}\right]_0^{T}.
\label{Stoy}\ea
In the context of the gravitational sequestering mechanism, $m(t)$ and $\lambda(t)$ correspond to the dynamical Planck mass squared and, respectively, the cosmological constant, $\eta(t)$ and $\rho(t)$ are the analogs of the 3-forms and their values are \emph{fixed at the boundary}\footnote{Note that this assumption does not spoil the well-posedness of the variational principle since one can still impose Dirichlet boundary conditions on the variation of $\eta$ and $\rho$.}, $q(t)$ encodes the metric degrees of freedom and $F(t)$ is an external force whose value is highly dependent on the effective description of its sources (in other words it is hard to compute analytically). The fact that $\eta(0)$, $\eta(T)$, $\rho(0)$ and $\rho(T)$ are all given {\it a priori} as boundary conditions is highly non-standard but it is the toy-model equivalent of specifying the value of the 4-form fluxes on the boundary of spacetime in the full model. The functions $f$ and $g$ are assumed to be regular, smooth and non-linear in $\lambda$ and $m$, respectively, and are the analogs of $\sigma$ and $\hat{\sigma}$ in \eqref{action2}. The boundary term can be understood as a Gibbons-Hawking-like term necessary for the well-posedness of the variational principle given the presence of second order time derivatives in the action. We will provide a simpler physical description of our toy model momentarily. 

The equations of motion are found by varying action (\ref{Stoy}). The variation with respect to $\eta$ and $\rho$ imposes that the $\lambda$ and $m$ are constants of motion
\be
\dot{\lambda}=\dot{m}=0\;.  
\label{constantsofmotion}\ee
Variation with respect to $\lambda$, $m$ and $q$ yields 
\ba
-q+\frac{1}{\lambda_{0}}f'\dot{\eta}=0\;,\label{eomlambda}\\
-\frac{1}{2}q\ddot{q}+\frac{1}{m_{0}}g'\dot{\rho}=0\;,\label{eommu}\\
-m\ddot{q}-\dot{m}\dot{q}-\frac{1}{2}\ddot{m}q-(\lambda+F)=0\;.\label{eomq1}
\ea
Using the constancy of $m$, the latter equation reduces to
\be
-m\ddot{q}-(\lambda+F)=0\label{eomq}\,.
\ee
We are now in a position where we can see that our toy model exhibits the sequestering-like behavior. Indeed suppose the force $F(t)$ contains a constant term $F_\text{div}$ that is formally divergent and that would infinitely backreact on the $q$ degree of freedom: $F=F_\text{div}+\bar{F}$. Integrating Eqs. \eqref{eomlambda} and \eqref{eommu}  and using the constancy of $\lambda$ and $m$, we get
\ba
\int_0^Tdt\, q=\frac{1}{\lambda_{0}}f'\left(\eta(T)-\eta(0)\right)\,,\label{inteom1}\\
\int_0^Tdt\, q\ddot{q}=2\frac{1}{m_{0}}g'\left(\rho(T)-\rho(0)\right)\,.\label{inteom2}
\ea
Multiplying Eq. \eqref{eomq} by $q$ and integrating then yields
\be
-m\int_0^T dt q\ddot{q}-\lambda\int_0^Tdt q - \int_0^T dtqF =0\,,  
\ee
which, upon using \eqref{inteom1} and \eqref{inteom2}, reduces to
\be
-\lambda=\langle F\rangle+\frac{2m\lambda_{0}g'}{m_{0}f'}\frac{(\rho(T)-\rho(0))}{(\eta(T)-\eta(0))}\,.
\label{lambda1}\ee
The first term in equation \eqref{lambda1} is the historic average of the source $F(t)$ {i\it .e.} $\langle F\rangle \equiv\int_0^T dtqF/\int_0^Tdt q$. It is easy to see that $\langle F\rangle=F_\text{div}+\langle \bar{F}\rangle$ and therefore Eq. \eqref{eomq} reduces to
\be
m\ddot{q}=-(\bar{F}+\lambda_\text{res})\,,
\label{eomq2}\ee
where $\lambda_\text{res}$ is a residual constant whose value only depends on the fixed boundary values of $\rho$ and $\eta$ and which is given by 
\be
-\lambda_\text{res}=\langle \bar{F}\rangle+\frac{2m\lambda_{0}g'}{m_{0}f'}\frac{\rho(T)-\rho(0)}{\eta(T)-\eta(0)}\,.
\label{lambdares}\ee
We thus see that the problematic term $F_\text{div}$ has completely dropped out of the equation of motion for $q$ having been replaced by a finite residual value. 

Now that we have found a simpler model exhibiting many of the features of vacuum energy sequestering, let us provide a physical interpretation. Consider the one dimensional problem of a charged point particle of mass $m$ and charge $e$ moving in a uniform electric field. We call $z$ the spatial dimension. Suppose the uniform field is created by a uniformly charged plane (of surface charge density $\sigma$) situated at $z=0$ and extending indefinitely in the orthogonal $x$ and $y$ dimensions. Suppose further that the particle is also feeling the effects of a uniform force whose value is extremely sensitive to the microscopic details of its source and which is formally divergent. The action for this model can be written as
\beq 
S=\int^{T}_{0}dt\left(\frac{1}{2}m\dot{z}^{2}-(\lambda+F)z\right)\,,
\label{toycharge}
\eeq
where $\lambda=e\sigma/2\epsilon_0$ and $F$ represents the UV divergent force.

To separate out and account for the divergence, we modify the action (\ref{toycharge}) by upgrading the charge $e$ and mass of the particle $m$ to the status of dynamical variables (for our purpose and to make the analogy with the toy model more apparent, we can choose units where $\sigma/2\epsilon_0=1$ so that $\lambda=e$). We then ensure the newly promoted variables $e$ and $m$ are constants of motion by adding ``gauge-fixing" terms $\eta$ and $\rho$. The modification we describe is equivalent to our toy model (\ref{Stoy}). A Lagrangian analysis would lead us to a global condition on $e$ and $m$ analogous to the one in (\ref{lambda1}). This acts like a constraint on the bare quantities $m$ and $e$, and therefore leads to the renormalization of the mass and charge of the particle. We also point out that residual (renormalized) values $e$ and $m$ are not determined via global conditions such as \eqref{lambda1}, but rather they are measured quantities. We will come back to this point in the next subsection.

On shell our toy model describes a particle moving in a one-dimensional renormalized linear potential. We will now proceed to do a Hamiltonian analysis of this model in order to exactly determine its physical degree of freedom count and obtain further insight into the nature of the global condition (\ref{lambda1}). 

\subsection{Hamiltonian and Hamilton's Equations}
\noindent

Here we reexamine our toy model from a Hamiltonian point of view and discuss its peculiar constraint structure. We first point out that our Lagrangian \eqref{Stoy} may be written as
\beq 
L=\hspace{-1mm}\frac{1}{2}m\dot{q}^{2}+\frac{1}{2}\dot{m}q\dot{q}-(\lambda+F)q+f\left(\frac{\lambda}{\lambda_{0}}\right)\dot{\eta}+g\left(\frac{m}{m_{0}}\right)\dot{\rho}\;.
\eeq

The canonical momenta are then given by 
\ba
p&=&\frac{\partial L}{\partial \dot{q}}=m\dot{q}+\frac{1}{2}\dot{m}q\;,\quad \pi_{m}=\frac{\partial L}{\partial\dot{m}}=\frac{1}{2}q\dot{q}\;,\label{momentatoy}\\
\pi_{\eta}&=&\frac{\partial L}{\partial\dot{\eta}}=f(\lambda)\;,\quad \pi_{\rho}=\frac{\partial L}{\partial\dot{\rho}}=g(m)\;,\quad \pi_{\lambda}=\frac{\partial L}{\partial\dot{\lambda}}=0\;.\nn
\ea
We immediately notice that these equations do not allow us to solve for all the velocities in terms of the canonical momenta, which means we are in the presence of a constrained system. Indeed, only $\dot{m}$ and $\dot{q}$ can be solved for:
\beq 
\dot{m}=\frac{2}{q}\left(p-\frac{2m\pi_{m}}{q}\right)\;,\quad \dot{q}=\frac{2\pi_{m}}{q}\;. 
\eeq
We therefore follow the usual Dirac procedure \cite{Henneaux:1992ig,Cianfrani:2008hz} for dealing with constrained Hamiltonian systems by treating the remaining velocities as Lagrange multipliers $\ell_{\eta}$, $\ell_{\rho}$ and $\ell_\lambda$. The Hamiltonian for our toy model then reads
\ba
H&=&p\dot{q}+\pi_{m}\dot{m}+\ell_{\lambda}\pi_{\lambda}+\ell_\eta\pi_\eta+\ell_{\rho}\pi_{\rho}-L\nn\\
&=&\frac{2\pi_{m}}{q}\left(p-\frac{m\pi_{m}}{q}\right)+(\lambda+F)q+\ell_{\lambda}\pi_{\lambda}+\ell_{\eta}(\pi_{\eta}-f)+\ell_{\rho}(\pi_{\rho}-g)\;,\label{Htoy}
\label{Hamtoy}\ea
and Hamilton's equations follow immediately
\ba
&\dot{\lambda}&=\ell_{\lambda}\;,\quad\quad\dot{m}=\frac{2}{q}\left(p-\frac{2m\pi_{m}}{q}\right)\;,\nn\\
&\dot{\pi}_{\lambda}&=-q+\frac{1}{\lambda_{0}}f'\ell_{\eta}\;,\quad\quad\dot{\pi}_{m}=\frac{2\pi^{2}_{m}}{q^{2}}\;,\nn\\
&\dot{p}&=\frac{2\pi_{m}}{q^{2}}\left(p-\frac{2m\pi_{m}}{q}\right)-(\lambda+F)\;,\quad \dot{q}=\frac{2\pi_{m}}{q} \;,\nn\\
&\dot{\rho}&=\ell_{\rho}\;,\quad\quad\dot{\eta}=\ell_{\eta}\;,\quad\quad\dot{\pi}_{\rho}=\dot{\pi}_{\eta}=0\;.
\label{Hamseqns}
\ea
Naturally, combining Eqs. (\ref{Hamseqns}) leads to the Euler-Lagrange equations  (\ref{constantsofmotion})--(\ref{eomq}). 
However since this is a constrained Hamiltonian system not all of these equations are independent.
To count the number of physical degrees of freedom of our system and interpret the global condition (\ref{lambda1}), we must determine its exact constraint structure in a systematic manner. We perform this analysis in the next subsection.

\subsection{Constraint structure}
\noindent

For convenience, we divide our Hamiltonian (\ref{Htoy}) into a \emph{canonical} contribution and a term which depends on constraints:
\be
H=H_{C}+\ell_{i}\varphi_{i}\;,
\ee
where
\ba
H_{C}&\equiv& \frac{2\pi_{m}}{q}\left(p-\frac{m\pi_{m}}{q}\right)+(\lambda+F)q\;,\\
\ell_{i}\varphi_{i}&\equiv& \ell_{\lambda}\pi_{\lambda}+\ell_{\eta}(\pi_{\eta}-f)+\ell_{\rho}(\pi_{\rho}-g)\;,
\label{HtoyCandconst}
\ea
and $i$ runs over the set $\{\lambda,\eta,\rho\}$.
Each of the constraints $\varphi_{i}$ is \emph{primary}, {\it i.e.}, it appears explicitly in the Hamiltonian.
Let us then check to see whether any of our constraints give rise to secondary constraints {\it i.e.}, whether their conservation during time evolution further restricts the dimension of phase space\footnote{By ``phase space'' we mean the space of physical motions \cite{Henneaux:1992ig}, i.e., the constraint manifold.}. Defining the Poisson bracket of two functions $F$, $G$ of the canonical variables $q_i$, $p_i$ by
\be
\{F,G\}\equiv\sum_i \left(\frac{\pa F}{\pa q_i}\frac{\pa G}{\pa p_i}-\frac{\pa F}{\pa p_i}\frac{\pa G}{\pa q_i}\right)\,,
\ee
the conservation of $\varphi_\lambda$ leads to
\ba
\dot{\varphi}_\lambda&=&\{\varphi_{\lambda},H\}
=-q+\frac{1}{\lambda_{0}}f'\ell_{\eta}\approx 0\;,
\label{constlambda}
\ea
where the {\it weak equality} symbol $\approx$ denotes equality up to a linear combination of the constraints. Similarly, conservation of $\varphi_\eta$ and $\varphi_\rho$ lead to 
\ba
\dot{\varphi}_\eta&=&\{\varphi_{\eta},H\}
=-\frac{1}{\lambda_{0}}f'\ell_{\lambda}\approx0\;,
\label{consteta}
\ea
and
\ba
\dot{\varphi}_\rho&=&\{\varphi_{\rho},H\}
=\frac{2}{m_{0}q^{2}}g'\left(2\pi_{m}m-pq\right)\approx0\;,
\ea
respectively. This shows that only $\varphi_{\rho}$ gives rise to a {\it secondary constraint}
\beq
\chi\equiv pq-2m\pi_{m}\approx0\;, \label{secondconst}
\eeq
since \eqref{constlambda} and \eqref{consteta} seemingly determine the Lagrange multipliers $\ell_\lambda$ and $\ell_\eta$ in terms of the canonical variables (but we will come back to this point below). Notice that Eqs. \eqref{consteta} and \eqref{secondconst} in conjunction with the first two equations in \eqref{Hamseqns} immediately imply that $\dot{\lambda}=\dot{m}=0$.

We now have to check whether $\chi$ gives rise to a tertiary constraint. It turns out it does not since 
\beq
\dot{\chi}=\left\{\chi,H\right\}=-(\lambda+F)q-\frac{2m}{m_{0}}g'\ell_{\rho}\approx0\;,
\label{constchi}
\eeq
and this relation determines $\ell_\rho$ in terms of the canonical variables.  Altogether we have three primary constraints $\varphi_{\lambda}$, $\varphi_{\eta}$, $\varphi_{\rho}$ and one secondary constraint $\chi$.
Note that all of these constraints are {\it second class}\footnote{Recall that a constraint is said to be {\it first class} if its Poisson brackets with all of the other constraints vanish weakly. It is said to be second class if it is not first class.}.

This is however not the whole story. Despite the fact that Eqs. \eqref{constlambda} and \eqref{constchi} are not constraints {\it per se}, they do not determine the Lagrange multipliers $\ell_\eta$ and $\ell_\rho$ completely. Indeed, since the functions $\eta$ and $\rho$ are given fixed values at the boundary, $\int dt\ell_\eta=\eta(T)-\eta(0)$ and $\int dt\ell_\rho=\rho(T)-\rho(0)$ are already predetermined. Therefore, keeping in mind that $\lambda$ and $m$ are constants of motion, integrating Eqs. \eqref{constlambda} and \eqref{constchi} yields
\ba
\frac{1}{\lambda_{0}}f'(\eta(T)-\eta(0))&\approx&\int_0^T dt q\,,\label{teleo1}\\
-\frac{2m}{m_{0}}g'(\rho(T)-\rho(0))&\approx&\lambda\int_0^T dt q+\int_0^T dt F\,.\label{teleo2}
\ea
These two truly global conditions can be regarded as constraining the conserved quantities $\lambda$ and $m$ teleologically\footnote{At least as long as $\eta(T)\neq\eta(0)$ and $\rho(T)\neq\rho(0)$. We restrict ourselves to this very generic case.}, in the sense that a phase space trajectory that doesn't saturate them is forbidden post-factum. In the following we will call such global consistency conditions {\it teleological constraints} since they do reduce the dimension of phase space by relating different conserved quantities to each other, albeit not in the usual way (wherein constraints reduce the number of initial conditions one can independently set). We emphasize here that our so-called teleological constraints are not constraints in the usual Hamiltonian sense, but they are rather a consequence of the non-standard (final) boundary conditions imposed on the degrees of freedom $\eta$ and $\rho$. Although these teleological constraints do not set the values of $\lambda$ or $m$ they may be understood as enforcing a renormalization condition as in Eq. (\ref{lambdares}).

We are now in a position to count the physical degrees of freedom of our system. Naively, we have twelve phase space degrees of freedom for this system, coming from the following canonical pairs of phase space coordinates: $(q,p), (m,\pi_{m}), (\eta,\pi_{\eta}), (\rho,\pi_{\rho})$ and $(\lambda,\pi_{\lambda})$. 
As we have seen above, there are however a certain number of constraints that reduce the dimension of phase space. The true number of physical degrees of freedom $N_{\rm phys}$ is usually given by
\beq 
N_{\rm phys}=N_{T}-N_{SC}-2N_{FC}\;,
\label{Nphys}\eeq
where $N_{T}$ denotes the total naive number of phase space degrees of freedom, $N_{SC}$ the number of second class constraints, and $N_{FC}$ the number of first class constraints. For our toy model we have $N_{T}=10$, $N_{SC}=4$, and $N_{FC}=0$. Therefore, 
\beq 
N_{\rm phys}=6\;.
\eeq
In summary, we are left with six propagating degrees of freedom: two coming from the particle motion (corresponding to the canonical pair $(q,p)$), one for each of the bare $\lambda$ and $m$, as well as two more corresponding to the functions $\eta$ and $\rho$. However $\lambda$ and $m$ are constants of motion on-shell and, moreover, the teleological constraints \eqref{teleo1} and \eqref{teleo2} completely restrict their possible values. As for $\eta$ and $\rho$, their values are fixed at the boundary from the get go so they do not have any dynamics. We are therefore left with only two degrees of freedom corresponding to the motion of a massive particle in a renormalized linear potential.

Let us compare the Hamiltonian analysis we used to study our toy model to a more traditional one to emphasize the differences and the importance of introducing final boundary conditions. The key difference between the two is that our analysis makes use of a non-standard initial value formulation of the problem. In a conventional initial value formulation we could have in principle solved for the auxiliary degrees of freedom $\eta$ and $\rho$ completely using Eqs. (\ref{eomlambda}) -- (\ref{eomq1}), dropping the final boundary conditions $\eta(T)$ and $\rho(T)$. In doing so, however, the solution will necessarily depend on $F_{\text{div}}$ such that $\lambda_{\text{res}}$ in \eqref{lambdares} is radiatively unstable. On the contrary, if we instead treat $\eta(T)$ and $\rho(T)$ as given final boundary conditions, we cannot use Eqs. (\ref{eomlambda}) -- (\ref{eomq1}) to solve for $\eta$ and $\rho$ completely, and, consequently, their integrated forms (\ref{inteom1}) and (\ref{inteom2}) give global relationships between $q$, $\lambda$, and $m$. Crucially, these global relations make the ratio in $\lambda_{\text{res}}$ (\ref{lambdares}) radiatively stable. We point out that introducing the final boundary conditions $\eta(T)$ and $\rho(T)$ is in the spirit of the original sequester model \cite{Kaloper:2013zca,Kaloper:2014dqa,Kaloper:2014fca}: neither $\eta$ nor $\rho$ are observable, while $\lambda_{\text{res}}$ is to be measured.

Now that we have worked out the Hamiltonian structure of this simple toy model of the vacuum energy sequestering mechanism, let us move on and extend this analysis to the gravitional action (\ref{action2}). 

\section{Hamiltonian Analysis of Local Vacuum Energy Sequestering}
\label{secHamseq}
\noindent

To do the Hamiltonian analysis of the sequestering model (\ref{action2}), we proceed in the usual way by performing an ADM split of spacetime, constructing the total Hamiltonian and studying its constraint structure. Since we have promoted $\kappa^{2}$ to a field $\kappa^{2}(x)$, the sequestering action shares similarities with Brans-Dicke theories of gravity, and therefore we may perform an analysis analogous to, e.g., \cite{Olmo:2011fh}. Notice that our analysis proceeds from a slightly different starting point then the one in \cite{Bufalo:2016omb} but leads to equivalent results.

The ADM split is as follows. The four dimensional spacetime manifold $\mathcal{M}$ is foliated by spacelike hypersurfaces $\Sigma_{t}$ which define the time coordinate $t$ and are endowed with spatial coordinates $y^i$, allowing us to determine the canonical momenta. Recall that \cite{Poisson:2009pwt}
\ba
g_{00}&=&-N^{2}+N^{i}N^{j}h_{ij}\;,\quad g_{0i}=N^{k}h_{ki}\,,\quad g_{ij}=h_{ij}\;,\nn\\
g^{00}&=&-\frac{1}{N^{2}}\;,\quad g^{0i}=\frac{N^{i}}{N^{2}}\;,\quad g^{ij}=h^{ij}-\frac{N^{i}N^{j}}{N^{2}}\;,\nn\\
\sqrt{-g}&=&N\sqrt{h}\;,\quad K_{ij}=\frac{1}{2N}(\dot{h}_{ij}-2D_{(i}N_{j)})\,,
\ea
where $h_{ij}$ is the three metric living on the constant time-slices $\Sigma_{t}$, $K_{ij}$ is its associated extrinsic curvature, $D_{i}$ the covariant derivative with respect to $h_{ij}$\footnote{When acting on a weight $w$ even tensor density $t^{ijk\dots}$, the covariant derivative $D_lt^{ijk\dots}$ is defined by $D_lt^{ijk\dots}\equiv \sqrt{h}^wD_l\left(\sqrt{h}^{-w} t^{ijk\dots}\right)$.} and $N$, $N^i$ are the lapse and shift, respectively. The latter are related to the time flow vector $t^{a}$ by $t^{a}=Nn^{a}+N^{a}$, where $n^{a}$ is a normalized timelike vector normal to $\Sigma_{t}$. Note that we will follow the usual custom of raising and lowering indices from the middle of the alphabet with the three-metric $h_{ij}$.

For concreteness we will restrict the matter sector to be a massive scalar field and thus we take
\beq 
\mathcal{L}_{m}(g^{\mu\nu},\phi)=-\frac{1}{2}g^{\mu\nu}\partial_{\mu}\phi\partial_{\nu}\phi-V(\phi)\;.
\eeq
We can now recast action \eqref{action2} in a form that is more conducive to a Hamiltonian analysis {\it i.e.}
\be 
S=\int dt\,L=\int dt\int d^3y\,\mathcal{L}\,,
\ee
where
\be
{\cal L}\equiv{\cal L}_{\rm grav}+{\cal L}_{m}+{\cal L}_{\tau}+{\cal L}_{\hat{\tau}}\,,
\ee
and
\ba
\mathcal{L}_{\rm grav}&=&N\sqrt{h}\biggr[\frac{\kappa^{2}}{2}(R^{(3)}+K_{ij}K^{ij}-K^{2})-\Lambda +\frac{h^{ij}}{N}(D_{i}N)(D_{j}\kappa^{2})\nn\\
&&-\frac{K}{N}\left(\dot{\kappa^{2}}-N^{i}D_{i}\kappa^{2}\right)\biggr]\,,
\label{LG}\\
 {\cal L}_{m}&=&N\sqrt{h}\,\bigg[\frac{1}{2N^{2}}\dot{\phi}^{2}-\frac{N^{i}}{N^{2}}\dot{\phi}\partial_{i}\phi-\frac{1}{2}\left(h^{ij}-\frac{N^{i}N^{j}}{N^{2}}\right)(\partial_{i}\phi)(\partial_{j}\phi)-V(\phi)\bigg]\,, \label{LM}\\
{\cal L}_{\tau}&=&\sigma\left(\frac{\Lambda}{\mu^{4}}\right)(\dot{\tau}+\partial_{i}\tau^{i})\,, \label{Ltau}\\
{\cal L}_{\hat{\tau}}&=&\hat{\sigma}\left(\frac{\kappa^{2}}{M^{2}_{P}}\right)(\dot{\hat{\tau}}+\partial_{i}\hat{\tau}^{i})\,.
\label{Ltauhat}
\ea
Here and henceforth a dotted quantity refers to the directional derivative with respect to the ADM time. In the above expressions, $\Lambda$ and $\kappa$ are local scalar fields, $\tau^{\mu}$ and $\hat{\tau}^{\mu}$ are vector densities related to the 3-forms $A$ and $\hat{A}$ by $\tau^{\mu}=4\epsilon^{\mu\nu\rho\sigma}A_{\nu\rho\sigma}$ and $\hat{\tau}^{\mu}=4\epsilon^{\mu\nu\rho\sigma}\hat{A}_{\nu\rho\sigma}$. We should also mention that the Gibbons-Hawking-like term required in order to have a well-defined variational principle at the boundary of spacetime yields a residual spacelike boundary term
\be
-\int_{\pa\Sigma_t} d^2s\,\kappa^2 kN\,,
\label{GBY}
\ee 
where $d^2s$ is the surface element on $\pa\Sigma_t$ and $k$ its extrinsic curvature. 

We can now see that the spatial metric conjugate momentum is
\ba
\pi^{ij}\equiv\frac{\delta L}{\delta \dot{h}_{ij}}=\frac{\pa\mathcal{L}}{\pa\dot{h}_{ij}}&=&\frac{\kappa^{2}}{2}\sqrt{h}(K^{ij}-h^{ij}K)-\frac{\sqrt{h}h^{ij}}{2N}[\dot{\kappa^{2}}-N^{l}D_{l}\kappa^{2}]\,,
\label{piij}\ea
where the 
functional derivative is generically defined for any functional $\mathcal{F}(f)$ by the relation
\be
\delta F(f)\equiv\int dy \frac{\delta \mathcal{F}(f)}{\delta f(y)}\delta f(y).
\label{funcderiv}\ee
Similarly we would get
\ba
\pi_{\kappa^{2}}&=&-\sqrt{h}K,\\
\pi_{\phi}&=&N\sqrt{h}\left(\frac{\dot{\phi}}{N^{2}}-\frac{N^{i}}{N^{2}}\partial_{i}\phi\right),\\
\pi_{\tau^{0}}&=&\sigma\left(\frac{\Lambda}{\mu^{4}}\right),\quad\quad\pi_{\hat{\tau}^{0}}=\hat{\sigma}\left(\frac{\kappa^{2}}{M^{2}_{P}}\right).
\label{canonicalmomenta}
\ea
Notice that while we cannot solve these equations for all the velocities in terms of the canonical momenta,
we do however obtain expressions for $\dot{h}_{ij}$, $\dot{\kappa^{2}}$ and $\dot{\phi}$ {\it viz.}
\ba
\dot{\phi}&=&\frac{N}{\sqrt{h}}\pi_{\phi}+N^{i}\partial_{i}\phi\;,\label{phidot}\\
\dot{\kappa^{2}}&=&N^{l}D_{l}\kappa^{2}-\frac{2N}{3\sqrt{h}}\chi\;, \label{kappadot}\\
\dot{h}_{ij}&=&\frac{4N}{\kappa^{2}\sqrt{h}}\biggr\{G_{ijkl}\pi^{kl}+\frac{1}{6}h_{ij}\chi\biggr\}+2D_{(i}N_{j)}\;,\label{hdot}
\ea
where we have introduced the so-called \emph{super-metric} $G_{ijkl}\equiv G_{klij}=h_{ik}h_{jl}-1/2h_{ij}h_{kl}$ and the quantity 
\be
\chi\equiv h_{ij}\pi^{ij}-\kappa^{2}\pi_{\kappa^{2}}\,.
\label{chiconstraint}
\ee 

We can now define the Hamiltonian density by analogy with our toy model Hamiltonian (\ref{Hamtoy}) by treating every unsolved for velocity as a Lagrange multiplier:
\ba
\mathcal{H}&\equiv&\pi^{ij}\dot{h}_{ij}+\pi_{\phi}\dot{\phi}+\pi_{\kappa^{2}}\dot{\kappa^{2}}-\mathcal{L} \nn\\
&&+\ell_{N}\pi_{N}+\ell_{N^{i}}\pi_{N^{i}}+\ell_{\Lambda}\pi_{\Lambda}+\ell_{\tau^{i}}\pi_{\tau^{i}}+\ell_{\hat{\tau}^{i}}\pi_{\hat{\tau}^{i}}+\ell_{\tau^{0}}\pi_{\tau^{0}}+\ell_{\hat{\tau}^{0}}\pi_{\hat{\tau}^{0}}\,.
\label{hamdens1}
\ea
Substituting the Lagrangian densities (\ref{LG})--(\ref{Ltauhat}) and the velocities (\ref{phidot})--(\ref{hdot}) into the Hamiltonian density (\ref{hamdens1}), and performing an integration by parts, we are led to
\ba
\mathcal{H}&=&N\mathcal{H}_0+N^{i}\mathcal{H}_i +\ell_{\Lambda}\pi_{\Lambda}+\mathcal{C}_{i}\tau^{i}+\hat{\mathcal{C}}_{i}\hat{\tau}^{i}\nn\\
&&+\ell_{\tau^{0}}\left[\pi_{\tau^{0}}-\sigma\left(\frac{\Lambda}{\mu^{4}}\right)\right]+\ell_{\hat{\tau}^{0}}\left[\pi_{\hat{\tau}^{0}}-\hat{\sigma}\left(\frac{\kappa^{2}}{M^{2}_{P}}\right)\right]+\ell_{N}\pi_{N}+\ell_{N^{i}}\pi_{N^{i}}\hspace{-1mm}+\ell_{\tau^{i}}\pi_{\tau^{i}}+\ell_{\hat{\tau}^{i}}\pi_{\hat{\tau}^{i}} \nn\\
&&+\text{boundary term}\;.
\label{totalhamiltonian}
\ea
Here we have introduced the \emph{super-Hamiltonian} $\mathcal{H}_0$ and \emph{super-momentum} $\mathcal{H}_{i}$,
\ba
\mathcal{H}_{0}&\equiv&\frac{2}{\kappa^{2}\sqrt{h}}G_{ijkl}\pi^{ij}\pi^{kl}-\sqrt{h}\left(\frac{\kappa^{2}}{2}R^{(3)}-\Lambda\right)+\frac{1}{3\kappa^{2}\sqrt{h}}\chi^{2}-D_{i}[h^{ij}\sqrt{h}D_{j}\kappa^{2}]\nn\\
&&+\sqrt{h}\left(\frac{\pi^{2}_{\phi}}{2h}+\frac{1}{2}h^{ij}(\partial_{i}\phi)(\partial_{j}\phi)+V(\phi)\right)\;,\label{superHG}\\
\mathcal{H}_{i}&\equiv&\pi_{\kappa^{2}}D_{i}\kappa^{2}-2h_{ij}D_{k}\pi^{jk}+\pi_{\phi}\partial_{i}\phi\;.\label{superPG}
\ea
and
\be
\mathcal{C}_{i}\equiv\pa_i\sigma\left(\frac{\Lambda}{\mu^{4}}\right)\,,\quad
\hat{\mathcal{C}}_{i}\equiv\pa_i\hat{\sigma}\left(\frac{\kappa^{2}}{M^{2}_{P}}\right)\,.
\ee

Note that Hamilton's equations resulting from the above Hamiltonian density, are consistent with the definition of the Lagrange multipliers {\it viz.}, $\ell_N=\dot{N}$, $\ell_{N^i}=\dot{N}^i$, $\ell_{\Lambda}=\dot{\Lambda}$, $\ell_{\tau^{0}}=\dot{\tau}^{0}$, $\ell_{\hat{\tau}^{0}}=\dot{\hat{\tau}}^{0}$, $\ell_{\tau^{i}}=\dot{\tau}^{i}$ and $\ell_{\hat{\tau}^{i}}=\dot{\hat{\tau}}^{i}$.

Lastly, the boundary terms in our total Hamiltonian density (\ref{totalhamiltonian}), explicitly given by
\ba
&&-\partial_{i}\left(\sigma\left(\frac{\Lambda}{\mu^{4}}\right)\tau^{i}\right)-\partial_{i}\left(\hat{\sigma}\left(\frac{\kappa^{2}}{M^{2}_{P}}\right)\hat{\tau}^{i}\right)+ D_{i}[h^{ij}N\sqrt{h}D_{j}\kappa^{2}+2\pi^{ij}h_{jk}N^{k}]\;,\label{boundarytermG}
\ea
ensure that the variational principle is well defined. More precisely, the integrations by parts required in order to write the Hamiltonian in ADM form introduce potentially dangerous terms such as first derivatives of the conjugate momenta or second derivatives of $\kappa^2$. The boundary terms~\eqref{boundarytermG} compensate for these artifacts.

With these definitions at hand, we can now proceed to determine the constraint structure of the theory. 

\subsection{Constraint Structure}
\noindent

We begin by integrating out $\ell_{N},\ell_{N^{i}},\ell_{\tau^{i}}$ and $\ell_{\hat{\tau}^{i}}$ to obtain the {\it reduced} Hamiltonian
\ba
\mathcal{H}'&=&N\mathcal{H}_0+N^{i}\mathcal{H}_{i}+\ell_{\Lambda}\pi_{\Lambda}+\mathcal{C}_{i}\tau^{i}+\hat{\mathcal{C}}_{i}\hat{\tau}^{i} \nn\\
&&+\ell_{\tau^{0}}\left[\pi_{\tau^{0}}-\sigma\left(\frac{\Lambda}{\mu^{4}}\right)\right]+\ell_{\hat{\tau}^{0}}\left[\pi_{\hat{\tau}^{0}}-\hat{\sigma}\left(\frac{\kappa^{2}}{M^{2}_{P}}\right)\right].
\label{totalhamiltonianred}
\ea
In this new set-up, $N, N_{i}, \tau_{i}$ and $\hat{\tau}_{i}$ are mere Lagrange multipliers\footnote{The fundamental reason we are able to do this reduction is that $\pi_{N}\approx\pi^{N^{i}}_{i}\approx\pi^{\tau^{i}}_{i}\approx\pi^{\hat{\tau}^{i}}_{i}\approx0$ are first class constraints, so each eliminate one canonical pair.}. We will take the reduced Hamiltonian (\ref{totalhamiltonianred}) to be the starting point of our constraint analysis.

It is easy to see that we have nine independent primary constraints. The first seven are
\ba
\mathcal{H}_0&\approx&\mathcal{H}_{i}\approx\pi_{\Lambda}\approx0\;,\nn\\
\varphi_{\tau^0}&\equiv&\pi_{\tau^{0}}-\sigma\left(\frac{\Lambda}{\mu^{4}}\right)\approx 0\;,\nn\\
\varphi_{\hat{\tau}^0}&\equiv&\pi_{\hat{\tau}^{0}}-\hat{\sigma}\left(\frac{\kappa^{2}}{M^{2}_{P}}\right)\approx 0\;.
\ea
The relationships $\mathcal{C}_{i}\approx\hat{\mathcal{C}_{i}}\approx0$ naively seem to give six additional primary constraints, however, only two of them are independent because of identities such as $\partial_{i}\mathcal{C}_{j}=\partial_{j}\mathcal{C}_{i}$, and $\partial_{i}\hat{\mathcal{C}}_{j}=\partial_{j}\hat{\mathcal{C}}_{i}$. Notice that these last two constraints are equivalent to $\pa_i\Lambda\approx\pa_i\kappa^2\approx 0$ and thus reduce $\Lambda$ and $\kappa^{2}$ to the status of global dynamical variables, i.e., independent of the spatial coordinates. 

We now check whether these primary constraints give rise to secondary constraints. Doing this involves calculating the Poisson brackets of the primary constraints with the Hamiltonian $H\equiv\int d^3y\,{\cal H}$. We do so by defining the Poisson bracket of two arbitrary functions of phase space $A$ and $B$ as
\be
\{A,B\}\equiv\sum_{i}\int d^{3}x\biggr(\frac{\delta A}{\delta q_{i}}\frac{\delta B}{\delta p_{i}}-\frac{\delta A}{\delta p_{i}}\frac{\delta B}{\delta q_{i}}\biggr)\;,
\ee
where the functional derivative has been defined in Eq. (\ref{funcderiv}).

The requirement that $\hat{\mathcal{C}}_{i}$ be conserved during time evolution {\it i.e.} 
\beq 
\dot{\hat{\mathcal{C}}}_{i}=\{\hat{\cal C}_{i},H\}=\frac{2N\hat{\sigma}'}{3M^{2}_{P}\sqrt{h}}\chi\approx0\,,
\eeq
where $\chi$ is defined in Eq. \eqref{chiconstraint}, leads to the secondary constraint
\beq
\chi= h_{ij}\pi^{ij}-\kappa^{2}\pi_{\kappa^{2}}\approx0\,.
\label{tertiary1}
\eeq
This turns out to be the only secondary constraint of the model. Note that (\ref{tertiary1}) and $\partial_{i}\kappa^{2}\approx0$ forces $\dot{\kappa^{2}}\approx0$ according to Eq. \eqref{kappadot}. Similarly $\dot{\phi}_{\tau^0}=\{\varphi_{\tau^0},H\}\approx 0$ implies that $\ell_\Lambda=\dot{\Lambda}\approx 0$.

The chain of constraints stops here since the Poisson bracket of the above constraints with the reduced Hamiltonian either is zero on the constraint surface or determines Lagrange multipliers in terms of dynamical variables. 

Just like in our toy model, however, not all relationships between dynamical variables and Lagrange multipliers arising from the evolution of the constraints determine the Lagrange multipliers completely. For example, $\dot{\pi}_\Lambda=\{\pi_{\Lambda},H\}\approx0$ gives
\beq 
N\sqrt{h}\approx\frac{1}{\mu^{4}}\sigma'\left(\frac{\Lambda}{\mu^{4}}\right)(\partial_{i}\tau^{i}+\ell_{\tau^{0}})\;,\label{cond1}
\eeq
while $\dot{\chi}=\{\chi,H\}$ yields
\ba 
\frac{N}{\kappa^2\sqrt{h}}G_{ijkl}\pi^{ij}\pi^{kl}&-&\frac{1}{4}N\sqrt{h}\kappa^2R^{(3)}-\frac{3}{2}N\sqrt{h}\Lambda
+\frac{1}{2}N\sqrt{h}\left(\frac{3\pi_\phi^2}{2h}-\frac{1}{2}h^{ij}\pa_i\phi\pa_j\phi-3V(\phi)\right)\nn\\
&\approx&\frac{\kappa^2}{M_P^2}\hat{\sigma}'\left(\frac{\kappa^2}{M_P^2}\right)(\partial_{i}\hat{\tau}^{i}+\ell_{\hat{\tau}^{0}})\,,
\label{cond2}
\ea
where we have used \eqref{tertiary1} to simplify the expressions. \eqref{cond1} and \eqref{cond2} are not constraints {\it per se}, since they could in principle be used to solve for the Lagrange multipliers $\ell_{\tau^{0}}$ and $\ell_{\hat{\tau}^{0}}$. However since $\ell_{\tau^{0}}=\dot{\tau}^0$ and $\ell_{\hat{\tau}^{0}}=\dot{\hat{\tau}}^0$, and the functions $\tau^{0}$, $\tau^i$ and $\hat{\tau}^{0}$, $\hat{\tau}^i$ are given fixed values at the boundary of spacetime, $\int dtd^{3}y\left(\pa_i\tau^i+\ell_{\tau^{0}}\right)$ and $\int dtd^{3}y\left(\pa_i\hat{\tau}^i+\ell_{\hat{\tau}^{0}}\right)$ are already predetermined. Therefore,  the integrated forms of (\ref{cond1}) and (\ref{cond2}) will lead to two new {\it  teleological constraints} just like in the toy model discussed in Sec.~\ref{sectoymodel}. Indeed using $\mathcal{H}_0\approx\mathcal{H}_{i}\approx0$, the constancy of $\Lambda$ and $\kappa^2$, and integrating over spacetime we arrive at
\ba
\int dtd^3y N\sqrt{h}&\approx&\frac{1}{\mu^{4}}\sigma'\int dt d^3y\left(\pa_i\tau^i+\ell_{\tau^{0}}\right)\,,\label{integralover}\\
\Lambda\int dtd^3y N\sqrt{h}&\approx&\frac{1}{4}\int dtd^3y\left(T_0{}^0+T_i{}^i\right)
-\frac{\kappa^{2}}{2M^{2}_{P}}\hat{\sigma}'\int dtd^{3}y(\partial_{i}\hat{\tau}^{i}+\ell_{\hat{\tau}^{0}})\,,
\label{integraloverbis}
\ea
where
\ba
T^{0}{}_{0}&\equiv&-\frac{1}{2h}\pi^{2}_{\phi}-\frac{1}{2}h^{ij}(\partial_{i}\phi)(\partial_{j}\phi)-V(\phi)-\frac{\pi_{\phi}N^{i}\partial_{i}\phi}{N\sqrt{h}}\,,\\
T^{i}{}_i&\equiv& \frac{3\pi_\phi^2}{2h}-\frac{1}{2}h^{ij}\pa_i\phi\pa_j\phi-3V(\phi)\,,
\ea
are related to the matter energy density and pressure as measured by a stationary observer. Note that  Eqs. (\ref{integralover}) and (\ref{integraloverbis}) only depend on gauge invariant global quantities and thus can really be considered as restricting the allowed trajectories of the system in phase space.

It is easy to see that by combining (\ref{integralover}) and (\ref{integraloverbis}) we also recover the global relation \eqref{lambdacond1}
\beq
\Lambda\approx\frac{1}{4}\langle T^\mu{}_\mu\rangle-\frac{\kappa^{2}\mu^{4}}{2M^{2}_{P}}\frac{\hat{\sigma}'}{\sigma'}\frac{\int dtd^{3}y(\partial_{i}\hat{\tau}^{i}+\ell_{\hat{\tau}^{0}})}{\int dtd^{3}y(\partial_{i}\tau^{i}+\ell_{\tau^{0}})}\,. \label{lambdacond2}
\eeq
As is the case with $\lambda_{\text{res}}$ in the toy model, the residual value of the cosmological constant will be radiatively stable because the functions $\tau^{0}$, $\tau^i$ and $\hat{\tau}^{0}$, $\hat{\tau}^i$ are specified on {\it the whole boundary of spacetime} and therefore the ratio on the right hand side \eqref{lambdacond2} has a definite value imposed by the boundary conditions (recall that $\ell_{\tau^0}=\dot{\tau}^0$ and the same is true for its hatted counterpart). A standard initial value formulation would have completely solved for $\tau^{0}$ and $\hat{\tau}^{0}$, which consequently would depend on the metric and vacuum fluctuations, thus rendering the aforementioned ratio and the residual cosmological constant UV sensitive.

\subsection{Counting Degrees of Freedom}
\noindent

Having worked out the system of constraints, let us now count the number of Hamiltonian degrees of freedom. Naively, the phase space degrees of freedom for our reduced  system are: (i) 12 degrees of freedom coming from $(h_{ij},\pi^{ij})$, (ii) 2 from the matter sector $(\phi,\pi_{\phi})$, (iii) 4 from the sector of fundamental constants, $(\kappa^{2},\pi_{\kappa^{2}})$ and $(\Lambda,\pi_{\Lambda})$, and (iv) the vector density sector degrees of freedom, specifically 2 from $(\tau^{0},\pi_{\tau^{0}})$, as well as an additional 2 from their hatted counterparts. 

Altogether this would correspond to 11 (local) canonical pairs, or a phase space dimension of $22$. However many of these degrees of freedom are auxiliary fields whose dynamics are completely set by the constraint equations computed in the previous section. More precisely, the primary constraints
\be
\pi_{\Lambda}\approx\mathcal{H}_0\approx\mathcal{H}_{i}\approx\varphi_{\tau^0}\approx\varphi_{\hat{\tau}^0}\approx0\,,
\ee
and secondary constraint
\beq 
\chi\approx0\;,
\eeq
each eliminate one degree of freedom {\it i.e.} 8. Recall however that the primary constraints
\be
\mathcal{C}_{i}\approx\hat{\mathcal{C}}_{i}\approx 0\,
\ee
only eliminate 2 degrees of freedom and moreover incompletely: they leave 2 global dynamical variables unconstrained. 

However, our model possesses three gauge symmetries: the familiar diffeomorphism invariance, as well as the two $A\rightarrow A+dB$, $\hat{A}\rightarrow\hat{A}+d\hat{B}$ invariances of the form sector. Therefore a subset of the above constraints must be {\it first class}. To see this we first shift the constraints $\mathcal{H}_0$, $\mathcal{C}_i$ and $\hat{\cal C}_i$ in the following way\footnote{We may determine the appropriate shifts by following the method outlined by Henneaux and Teitelboim in \cite{Henneaux:1992ig}, and exemplified in \cite{Bufalo:2016omb}. This involves solving for the Lagrange multipliers $\ell_{\tau^{0}}$ and $\ell_{\hat{\tau}^{0}}$ using our expressions for $\dot{\pi}_{\Lambda}$ (\ref{cond1}) and $\dot{\chi}$ (\ref{cond2}), and substituting back into the reduced Hamiltonian (\ref{totalhamiltonianred}).}:
\ba
\mathcal{H}_0'&\equiv&\mathcal{H}_0+\frac{\mu^4}{\sigma'}\varphi_{\tau^0}+\frac{M_P^2}{\kappa^2\hat{\sigma}'}\biggr[\frac{1}{\kappa^2\sqrt{h}}G_{ijkl}\pi^{ij}\pi^{kl}\nn\\
&&-\frac{1}{4}\sqrt{h}\kappa^2R^{(3)}-\frac{3}{2}\sqrt{h}\Lambda+\frac{1}{2}\sqrt{h}T^{i}{}_i\biggr]\varphi_{\hat{\tau}^0}\,,\label{shiftedconstraints1}\\
\mathcal{C}_i'&\equiv&\mathcal{C}_i+\pa_i\varphi_{\tau^0}=\pa_i\pi_{\tau^0}\,,\label{shiftedconstraints2}\\
\hat{\mathcal{C}}_i'&\equiv&\hat{\mathcal{C}}_i+\pa_i\varphi_{\hat{\tau}^0}=\pa_i\pi_{\hat{\tau}^0}\,.
\label{shiftedconstraints3}\ea 
It is then straightforward, albeit tedious, to check that $\mathcal{H}_0'$, $\mathcal{H}_i$, $\mathcal{C}_i'$ and $\hat{\mathcal{C}}_i'$ are first class constraints {\it i.e.}, that their Poisson brackets with all the constraints vanish on-shell\footnote{Specifically, the constraint algebra for our model is similar to that of general relativity, as presented in \cite{DeWitt:1967yk}, \emph{e.g.}, $\{\mathcal{H}_{0}'(y),\mathcal{H}_{i}(y')\}=\partial_{i}(\mathcal{H}_{0}'(y)\delta(y-y'))\approx0$. Meanwhile, $\chi$ has weakly vanishing Poisson brackets with (\ref{shiftedconstraints1})--(\ref{shiftedconstraints3}), \emph{e.g.}, $\{\chi(y),\mathcal{H}_{i}(y')\}=\partial_{i}(\chi(y)\delta(y-y'))\approx0$, but not with $\varphi_{\hat{\tau}^{0}}$, and it is therefore second class.}. Therefore $\mathcal{H}_0'$ and $\mathcal{H}_i$ actually eliminate 8 degrees of freedom instead of just 4, while $\mathcal{C}_i'$ and $\hat{\mathcal{C}}_i'$ eliminate 4 degrees of freedom incompletely and not just 2.

We are thus left with 6 local degrees of freedom (corresponding to the gravitational and matter propagating modes) and 4 global ones, namely $\Lambda$, $\kappa^2$, $\int d^3y\,\tau^0$ and $\int d^3y\,\hat{\tau}^0$. But $\tau^0$ and $\hat{\tau}^0$ have no dynamics since their boundary values are completely specified, while $\Lambda$ and $\kappa^{2}$ are constants on-shell and their values are set by the two global teleological constraints  (\ref{integralover}) and (\ref{integraloverbis}) (as long as the boundary terms are non-zero). This reduces the final number of physical degrees of freedom to $N_{\text{phys}}=6$.  In summary, we have six local propagating degrees of freedom, four coming from the gravitational field and two from the matter sector of a single real scalar field. This confirms that our theory propagates the same number of degrees of freedom as general relativity on-shell. However it enjoys the additional property that the Planck mass and cosmological constant obey the teleological constraint \eqref{lambdacond2} which enforces the degravitation of the vacuum energy loops. This is a consequence of the fact that the values of the 4-form fluxes are picked as boundary conditions for the model. It may also be a sign of the self-tuning properties of the theory \cite{Niedermann:2017cel}.

\section{Conclusion}
\label{secconc}
\noindent


In this article we provided a Hamiltonian analysis of a manifestly local and diffeomorphism invariant modification to general relativity which, under minimal assumptions, has been shown to remove the radiatively unstable contribution to the cosmological constant due to the vacuum energy generated by matter loops \cite{Kaloper:2013zca,Kaloper:2014dqa,Kaloper:2014fca,Kaloper:2015jra,Kaloper:2016yfa,DAmico:2017ngr}. To illustrate the mechanism of vacuum energy sequestering, we developed a toy model with most of the relevant features, and demonstrated that the removal of unphysical divergences in the vacuum energy emerges as a global condition as a consequence of what we termed teleological constraints. These global relations are a symptom of the non-standard boundary conditions imposed on the model wherein some degrees of freedom are constrained on the whole boundary of spacetime. This is the main observation of this work which builds upon previous Hamiltonian discussions of vacuum energy sequestering~\cite{Kluson:2014tma,Bufalo:2016omb}. The additional final boundary conditions are the key to understanding the sequestering mechanism and have not to our knowledge been previously discussed in the literature. Our Hamiltonian analysis of the gravitational model revealed that, unseen in the Lagrangian based analysis, these teleological constraints  conspire to degravitate the matter sector vacuum energy. In a standard initial value formulation, e.g., \cite{Bufalo:2016omb}, the auxiliary fields $F$ and $\hat{F}$ would be seen to depend on the metric and vacuum fluctuations, rendering their fluxes UV sensitive. It is only thanks to the teleological constraints that the residual cosmological constant can achieve radiative stability in the sequestering model. We also confirmed that the proposed model of vacuum energy sequestering is indistinguishable from general relativity on-shell. Note in particular that the evolution problem is well-posed in the gravitational sector despite the global nature of the residual cosmological constant.


The cancellation of the vacuum energy in the sequestering mechanism is enforced by two approximate symmetries of the theory \cite{Kaloper:2014dqa}: (i) a shift symmetry $\mathcal{L}_{m}\to\mathcal{L}_{m}+\nu^{4}$, $\Lambda\to\Lambda-\nu^{4}$, where $\nu$ is a constant, and (ii) a scaling symmetry of $\kappa^{2}$. While these two symmetries cancel the unphysical divergence coming from the matter sector loops, the fact that the scaling symmetry is broken at finite Planck mass $M_{\text{P}}$ implies that divergences arising from the graviton sector loops will not be cancelled in the sequestering theory described by (\ref{action2}). To deal with the graviton loops, it was proposed in \cite{Kaloper:2016jsd} to modify the action (\ref{action2})  by adding a Gauss-Bonnet contribution $R_{GB}=R^{2}-4R_{\mu\nu}^{2}+R^{2}_{\mu\nu\rho\sigma}$, which only alters the topological sector in four dimensional spacetime, leaving local phenomena unaffected. Then, by introducing a Gauss-Bonnet coupling $\theta$ and upgrading it to an auxiliary field $\theta(x)$, it was shown that this improved model can sequester all large contributions from graviton loops sourced by the Einstein equations. We believe it would be interesting to provide a Hamiltonian analysis of this modified sequestering model, and leave this to future work. 

Having performed a Hamiltonian analysis of the sequestering model, it is also natural to try to quantize the theory. This may be accomplished by either following the canonical approach \cite{DeWitt:1967yk} where Poisson brackets are upgraded to Dirac brackets, or using a Hamiltonian path integral analysis, as has been performed in \cite{Bufalo:2016omb}. Following the path integral techniques developed in \cite{Bufalo:2016omb}, it would be interesting to study how the sequestering model differs from general relativity on a quantum level, and see if this can shed light on the problem of time in quantum gravity, as attempted in models of unimodular gravity \cite{Unruh:1989db,Kuchar:1991xd}. We leave these questions for future study.

\hspace{2mm}

\acknowledgments

We  are  grateful  for  comments  by Nemanja Kaloper, Tony Padilla and Tanmay Vachaspati. GZ is supported by John Templeton
Foundation  grant  60253.   


\bibliography{hamiltonseq}

\end{document}